\documentclass[UTF-8,preprint,showpacs,preprintnumbers,superscriptaddress]{revtex4-1}
\pdfoutput=1
\usepackage{multirow}
\usepackage{diagbox}
\usepackage{booktabs}
\usepackage{array}
\usepackage{natbib}
\usepackage{amsmath,amssymb,graphicx,bm,subfigure,float,siunitx,color}
\usepackage{bm,subfigure,float,siunitx,graphicx}
\usepackage[utf8]{inputenc}
\usepackage[figuresright]{rotating}
\usepackage{color}
\usepackage{mathrsfs}
\begin{document}
\title{Reconstructing square-law k-inflation from Planck data}
\author{Ming Liu}
\affiliation{College of Physics Science and Technology, Hebei University, Baoding 071002, China}
\author{Tong-Yu He}
\affiliation{College of Physics Science and Technology, Hebei University, Baoding 071002, China}
\author{Bohai Chen}
\affiliation{School of Liberal Arts and Sciences, North China Institute of Aerospace Engineering, Langfang 065000, China}
\author{Zhan-Wen Han}
\affiliation{College of Physics Science and Technology, Hebei University, Baoding 071002, China}
\affiliation{Yunnan Observatories, Chinese Academy of Sciences, Kunming 650216, China}
\author{Rong-Jia Yang \footnote{Corresponding author}}
\email{yangrongjia@tsinghua.org.cn}
\affiliation{College of Physics Science and Technology, Hebei University, Baoding 071002, China}
\affiliation{Hebei Key Lab of Optic-Electronic Information and Materials, Hebei University, Baoding 071002, China}
\affiliation{National-Local Joint Engineering Laboratory of New Energy Photoelectric Devices, Hebei University, Baoding 071002, China}
\affiliation{Key Laboratory of High-pricision Computation and Application of Quantum Field Theory of Hebei Province, Hebei University, Baoding 071002, China}

\begin{abstract}
We explore a square-law k-inflation using the Hamilton-Jacobi approach. Focusing on scenarios where the Hubble parameter exhibits a power-law dependence on the k-field, our analysis encompasses the computations of crucial observables, such as the scalar power spectrum, the tensor-to-scalar ratio, and the scalar spectral index. We further constrain the model's parameters using Planck data and present a specific form of the potential. Our results demonstrate that the model aligns well with observational data.
\end{abstract}

\maketitle

\section{Introduction}
The inflationary paradigm holds promise in addressing key challenges of standard hot big bang cosmology, offering a mechanism for generating primary density perturbations \cite{Guth:1980zm, Linde:1981mu, Baumann:2009ds}. One of the central inquiries in modern cosmology revolves around unraveling the mechanism that initiates the inflationary epoch, leading to the rapid expansion of the early universe. Over recent years, a plethora of theoretical models has emerged to elucidate this phenomenon. Many successful inflationary scenarios feature a single scalar field, often referred to as the ``inflaton", dynamically traversing its potential. K-inflation, characterized by a non-standard kinetic term, has gained prominence as a compelling model driving inflationary evolution \cite{Armendariz-Picon:1999hyi}.

Scalar fields with non-canonical kinetic terms, a common occurrence in supergravity and superstring theories, offer a more natural satisfaction of the slow-roll conditions crucial for inflation. K-inflation, boasting a variety of plausible models, exhibits diverse non-canonical terms, often stemming from different methods of achieving compactification. Notably, the inclusion of non-canonical terms in k-inflation has a pronounced effect, substantially reducing the tensor-to-scalar ratio \cite{Barenboim:2007ii, Franche:2009gk, Unnikrishnan:2012zu}. The extensive study of k-inflation has yielded various models and approaches, exemplified by \cite{Armendariz-Picon:1999hyi, Garriga:1999vw, Odintsov:2021lum, Granda:2021xyc, Pareek:2021lxz, Martin:2013uma, Mikura:2021ldx, Mohammadi:2019qeu, Mohammadi:2018wfk, Barenboim:2007ii, Lola:2020lvk, Shumaylov:2021qje, Feng:2014pta, Franche:2009gk, Unnikrishnan:2012zu, Gwyn:2012ey, Easson_2013, Zhang:2014dja, Rezazadeh:2014fwa, Cespedes:2015jga, Li_2012, Kamenshchik:2018sig,Yang:2023rjt,Gao:2024ybz,FerreiraJunior:2023qxi,Ageeva:2023nwf}.

In addition to the commonly employed slow-rolling approximation, another valuable approach for investigating inflation is the Hamilton-Jacobi formalism \cite{Salopek:1990jq, Muslimov:1990be, Lidsey:1991zp, Lidsey:1995np}. In this alternative method, instead of introducing a potential, the Hubble parameter is expressed as a function of the scalar field. By adopting this strategy, the model parameters can be expressed in terms of the Hubble parameter and its first derivative. This methodology has been widely employed to explore diverse inflationary models, as evident in existing literature \cite{Liddle:1994dx, Kinney:1997ne, Guo:2003zf, Aghamohammadi:2014aca, Sheikhahmadi:2016wyz, Videla:2016ypa, Sayar:2017pam,Yang:2023rjt,Zhang:2025tpg}. Exploring k-inflation within the framework of the Hamilton-Jacobi formalism presents an intriguing avenue. We anticipate that this approach can provide valuable insights, allowing for the derivation of key parameters and the determination of free parameters based on the latest observational data. Our analysis aims to demonstrate that all relevant parameters can be readily derived, facilitating the determination of free parameters through the utilization of up-to-date observational data.

The paper is structured as follows: In Sec. II, we provide an overview of the general framework, covering topics such as k-essence cosmology, the Hamilton-Jacobi formalism, attractor behavior, and cosmological perturbations. Sec. III serves as an application section, where we specifically explore a scenario in which the Hubble parameter is modeled as a power-law function of the k-field. We delve into the observational constraints on this model. Finally, in section IV, we offer a succinct summary of the obtained results.

\section{General framework}
This section provides a concise review of k-essence cosmology, the Hamilton-Jacobi formalism, the attractor behavior, and the cosmological perturbations. We present fundamental equations essential for subsequent calculations.
\subsection{K-essence cosmology}
K-inflation, characterized by a scalar field with non-canonical kinetic terms which appear generically in the effective action in string and supergravity theories, minimally coupled with gravity is described by the following action \cite{Armendariz-Picon:2000nqq, Armendariz-Picon:2000ulo, Malquarti:2003nn}
\begin{eqnarray}
\label{1}
S=\int d^4 x\sqrt{-g}\left[-\frac{m_{\rm{p}}^{2}}{16\pi} R+\mathcal{L}_{\phi}\left(\phi, X\right)\right],
\end{eqnarray}
where $m_{\rm{p}}$ represents the Planck mass, $R$ is the Ricci scalar, $\phi$ denotes the scalar field with kinetic terms $X=\nabla_{\mu} \phi \nabla^{\mu} \phi/2$, and $\mathcal{L}_{\phi}$ is the Lagrangian density of the scalar field. It is important to note that we use natural units with $c=\hbar=1$. The field equations are obtained through variation of the action with respect to the scalar field
\begin{equation}
\frac{\partial \mathcal{L}}{\partial \phi}-\left(\frac{1}{\sqrt{-g}}\right) \partial_\mu\left[\sqrt{-g} \frac{\partial \mathcal{L}}{\partial\left(\partial_\mu \phi\right)}\right]=0.
\end{equation}
The components of the energy-momentum tensor for a scalar field in a homogeneous and isotropic universe yield the pressure and the energy density, respectively.
\begin{eqnarray}
\label{2}
p_{\phi}=\mathcal{L}_{\phi},~~~~~\rho_{\phi}=2X\mathcal{L}_{\phi,X}-\mathcal{L_\phi},
\end{eqnarray}
where $\mathcal{L}_{\phi,X}=\partial\mathcal{L}_{\phi}/\partial{X}$. The speed of sound for perturbations is defined as
\begin{eqnarray}
\label{ss}
c^2_{\rm{s}}=\frac{p_{,X}}{\rho_{,X}}=\frac{\mathcal{L}_{\phi,X}}{2X\mathcal{L}_{\phi,XX}+\mathcal{L}_{\phi,X}}.
\end{eqnarray}
Structuring a Lagrangian density for k-inflation with non-canonical kinetic terms, the usual way is to add or multiply the kinetic term in the canonical Lagrangian density with the power of kinetic term: $X^{1/2}$, $X$, $X^{3/2}$, and so on. Here we consider the following simple case
\begin{eqnarray}
\label{4}
\mathcal{L}_{\phi}=\frac{\alpha\left(2X\right)^{2}}{m^4_{\rm{p}}}-V(\phi).
\end{eqnarray}
This k-inflation model was discussed in \cite{Li_2012} by employing the slow-rolling approximation. Here we reconsider it by using the Hamilton-Jacobi approach. It will be interesting and valuable to compare the results obtained from these two schemes. The viability of the theory hinges on specific conditions, as demonstrated in \cite{Bruneton:2007si}. For a well-defined theory, it is required that $\mathcal{L}_{\phi,X} > 0$ and $2X\mathcal{L}_{\phi,XX} + \mathcal{L}_{\phi,X} > 0$, leading to the constraint $\alpha > 0$. By utilizing Eqs. (\ref{2}) and (\ref{4}), we obtain the associated pressure and energy density, respectively, as
\begin{eqnarray}
\label{5}
p_\phi=\frac{4 a X^2}{m^4_{p}}-V(\phi),
\end{eqnarray}
\begin{eqnarray}
\label{6}
\rho_\phi=\frac{12 a X^2}{m^4_{p}}+V(\phi).
\end{eqnarray}
From Eq. (\ref{ss}), we find the speed of sound as: $c^2_{\rm{s}}=1/3$.

Recent observations strongly support the notion that the universe exhibits homogeneity, isotropy, and spatial flatness. This geometric configuration is aptly described by the Friedmann-Lemaître-Robertson-Walker (FLRW) metric
\begin{eqnarray}
\label{7}
d s^{2}=d t^{2}-a^{2}(t)\left[d r^{2}+r^{2}\left(d \theta^{2}+\sin ^{2} \theta d \varphi^{2}\right)\right].
\end{eqnarray}
Within this spacetime, where $X=\dot{\phi^2}/2$, the Friedmann equations for k-inflation (\ref{4}) take the forms
\begin{eqnarray}
\label{8}
H^{2}(\phi)=\frac{8\pi}{3m^{2}_{\rm{p}}}\rho_\phi=\frac{8 \pi \alpha \dot{\phi} ^4}{m^6_{\rm{p}}}+\frac{8 \pi  V}{3 m^2_{\rm{p}}},
\end{eqnarray}
\begin{eqnarray}
\label{9}
\dot{H}(\phi)=-\frac{4\pi}{m^{2}_{\rm{p}}}(\rho_\phi+p_\phi)=-\frac{16 \pi  \alpha \dot{\phi}^4}{m^6_{\rm{p}}}.
\end{eqnarray}
The field equation for the k-field, derived from Eq. (\ref{1}), is given by
\begin{eqnarray}
\label{3}
(\mathcal{L}_{\phi,X}+2X \mathcal{L}_{\phi,XX})\ddot{\phi}+3\mathcal{L}_{\phi,X} H\dot{\phi}+\mathcal{L}_{\phi,\phi}=0.
\end{eqnarray}
This equation can also be derived from the conservation equation.

\subsection{Hamilton-Jacobi Formalism}
In Hamilton-Jacobi Formalism \cite{Salopek:1990jq,Muslimov:1990be,Lidsey:1991zp,Lidsey:1995np}, the Hubble parameter  is expressed as a function of the scalar field, denoted as $H:=H(\phi)$. Consequently, the time variable of $H$ can be reformulated as $\dot{H}=\dot{\phi} H^{\prime}$, where the prime represents the derivative with respect to the scalar field. Utilizing Eq. \eqref{9}, the time derivative of the scalar field can be derived in terms of the scalar field as follows
\begin{eqnarray}
\label{dotphi}
\dot{\phi}=m^2_{\rm{p}}\left(-\frac{H^{\prime}}{16\pi\alpha}\right)^{1/3}.
\end{eqnarray}
By substituting Eq. (\ref{dotphi}) into the Friedmann equation (\ref{8}), we obtain
\begin{eqnarray}
\label{hj}
H^2+\frac{m^2_{\rm{p}}}{4}  \sqrt[3]{-\frac{H^{\prime 4}}{2 \pi  \alpha}}-\frac{8 \pi  V}{3 m^2_{\rm{p}}}=0.
\end{eqnarray}
Equation (\ref{hj}) is known as the Hamilton-Jacobi equation. The potential of the model can be easily obtained as a function of the scalar field from Eq. \eqref{8}.
\begin{eqnarray}
\label{potential}
V(\phi)=\frac{3m^4_{\rm{p}} }{32\pi }\left ( -\frac{H^{\prime 4}}{2\pi a}  \right )^\frac{1}{3} +\frac{+3m^2_{\rm{p}} H^2 }{8\pi }.
\end{eqnarray}
The precise shape of the potential is unknown and must be postulated. The slow-roll parameters are defined as \cite{Panotopoulos:2007ky}
\begin{eqnarray}
\label{slow}
\epsilon(\phi)=2 c_{s} m^2_{\rm{p}} \left(\frac{H^{\prime}}{H}\right)^2,~~~~\eta(\phi)=2 c_{s} m^2_{\rm{p}}\frac{ H^{\prime \prime}}{H}.
\end{eqnarray}
In the slow-roll approximation, the universe experiences a quasi-de Sitter expansion during inflation, and the slow-roll parameters must be much smaller than unity $\epsilon, |\eta|\ll 1$. When $\ddot{a}$ vanishes, or equivalently, the slow-roll parameter $\epsilon$ reaches unity, inflation comes to an end. Thus, at the conclusion of inflation, we have
\begin{eqnarray}
\label{slowend}
H=\sqrt{2c_{s}}m_{\rm{p}}H^{\prime}.
\end{eqnarray}
The parameter that quantifies the amount of expansion, known as the number of e-folds, is defined as
\begin{eqnarray}
\label{efold}
N \equiv \int_{t_{\rm{i}}}^{t_{\rm{e}}} H d t=\int_{\phi_{\rm{i}}}^{\phi_{\rm{e}}} \frac{H(\phi)}{\dot{\phi}} d \phi,
\end{eqnarray}
where the subscript ``i" and ``e" represent the beginning and the end of inflation, respectively. In the Hamilton-Jacobi approach, it appears that the main parameters of the model can be derived more straightforwardly than in the slow-rolling approach, with fewer assumptions.

\subsection{Attractor behavior}
In accordance with Ref. \cite{liddle2000cosmological}, considering a homogeneous perturbation $\delta H$ to a solution $H(\phi)$, we can straightforwardly assess whether all potential trajectories or solutions converge to a shared attractor solution using the Hamilton-Jacobi method. The attractor condition is deemed satisfied if the perturbation $\delta H$ diminishes over time. By inserting $H(\phi) + \delta H(\phi)$ into Eq. (\ref{hj}) and linearizing, we derive
\begin{eqnarray}
\label{attra}
H^{\prime 1 / 3} \delta H^{\prime}(\phi)-\frac{3 \cdot 2^{4 / 3} \alpha^{1 / 3} \pi^{1 / 3}}{m_{\mathrm{p}}^{2}} H(\phi) \delta H(\phi) \simeq 0,
\end{eqnarray}
Solving this equation yields
\begin{eqnarray}
\label{solattra}
\delta H(\phi)=\delta H\left(\phi_{\mathrm{i}}\right) \exp \left[\frac{3 \cdot 2^{4 / 3} \alpha^{1 / 3} \pi^{1 / 3}}{m_{\mathrm{p}}^{2}} \int_{\phi_{\mathrm{i}}}^{\phi} \frac{H(\phi)}{H^{\prime 1 / 3}(\phi)} d \phi\right],
\end{eqnarray}
Where $\delta H(\phi_{\rm{i}})$ is the initial value of the perturbation. Given $H(\phi)$, we can analyze the behavior of the perturbation $\delta H(\phi)$. So, it is easy to consider the attractor behavior of solutions using the Hamilton-Jacobi method.

Using (\ref{dotphi}) and (\ref{efold}), we find the following simple expression describing the decay of perturbations
\begin{eqnarray}
\label{sola}
\delta H(\phi)=\delta H(\phi_{\rm{i}})\exp(-3N),
\end{eqnarray}
The provided statement indicates a notably swift convergence towards the inflationary attractor solution, demonstrating an exponential trend. Notably, the expression (\ref{sola}) remains independent of the free parameter associated with our model (\ref{4}). This implies that homogeneous perturbations in the considered inflationary model decay in a manner precisely analogous to the decay observed for canonical scalars, as discussed in \cite{Salopek:1990jq}.

\subsection{Cosmological perturbations}
Examine linearized scalar and tensor perturbations within the framework of the spatially flat Friedmann-Lemaître-Robertson-Walker (FLRW) metric, which can be characterized by the following line element \cite{Mukhanov:1990me, Kodama:1984ziu, Bardeen:1980kt}
\begin{equation}
\mathrm{d} s^2=(1+2 A) \mathrm{d} t^2-2 a(t)\left(\partial_i B\right) \mathrm{d} t \mathrm{~d} x^i-a^2(t)\left[(1-2 \psi) \delta_{i j}+2\left(\partial_i \partial_j E\right)+h_{i j}\right] \mathrm{d} x^i \mathrm{~d} x^j,
\end{equation}
the symbols $A$, $B$, $E$, and $\psi$ represent scalar metric perturbations, while $h_{ij}$ characterizes tensor perturbations. The curvature perturbation $\mathcal{R}$ on the uniform field slicing is defined as a gauge-invariant combination of the scalar field perturbation $\delta\phi$ and the metric perturbation $\psi$
\begin{equation}
\label{R}
\mathcal{R} \equiv \psi+\left(\frac{H}{\dot{\phi}}\right) \delta \phi.
\end{equation}
Derived from the equation governing the evolution of perturbations in the scalar field and the linearized Einstein's equation $\delta G_{\mu\nu}=k\delta T_{\mu\nu}$, the equation (\ref{R}) can be expressed as
\begin{equation}
\label{rk}
\mathcal{\ddot{R}}_k+2\left(\frac{\dot{z}}{z}\right) \mathcal{\dot{R}}_k+c_s^2 k^2 \mathcal{R}_k=0,
\end{equation}
where the dot represents the derivative with respect to conformal time, $\eta=\int dt/a(t)$, and the variable $z$ is defined as $z \equiv \frac{a\left(\rho_\phi+p_\phi\right)^{1 / 2}}{c_s H}$. Expressing Eq. (\ref{rk}) in the terms of the Mukhanov-Sasaki variable and after a bit tedious calculations, the scalar and tensor power spectra in the slow-roll limit can be derived as follows \cite{Garriga:1999vw}
\begin{eqnarray}
\label{ps}
\mathcal{P}_S(k)=\left[\frac{H^2}{2 \pi \left[c_s\left(\rho_\phi+p_\phi\right)\right]^{1/2}}\right]^2,
\end{eqnarray}
and
\begin{equation}
\label{pt}
\mathcal{P}_T(k)=\frac{16}{\pi}\left(\frac{H}{m_{\rm{p}}}\right)^2.
\end{equation}
The model's parameters can be constrained by leveraging observational data related to the scalar power spectrum $\mathcal{P}_S(k)$.

\section{Application}
Up to this point, fundamental equations have been derived, yielding preliminary outcomes. In this section, we will explore a specific form for the parameter $H(\phi)$ in relation to the scalar field to obtain more detailed results. We assume that the Hubble parameter follows a power-law function of the scalar field, allowing for further specificity in our findings.
\begin{eqnarray}
\label{pow}
H(\phi)=\mathcal{H}_{1} \left(\frac{\phi}{m_{\rm{p}}}\right)^{n},
\end{eqnarray}
Here, $n$ and $\mathcal{H}_{1}$ are constants. To simplify, we assume $\mathcal{H}_{1}=\beta m_{\rm{p}}$, where $\beta$ is a dimensionless constant. Utilizing Eq. \eqref{dotphi}, we can express the time derivative of the scalar field as
\begin{eqnarray}
\label{17}
\dot{\phi}=m^{2}_{\rm{p}}\sqrt[3]{-\frac{\beta n  }{16 \pi  \alpha}}\left(\frac{\phi}{m_{\rm{p}}}\right)^\frac{n-1}{3}.
\end{eqnarray}
The general form for the potential can be deduced from Eq. \eqref{potential}.
\begin{eqnarray}
\label{V}
V(\phi)=\frac{3\beta^2m^4_{\rm{p}}}{8\pi }\left ( \frac{\phi }{m_{\rm{p}}}  \right )^{2n } -\frac{3\beta^\frac{4}{3}n^\frac{4}{3}m^4_{\rm{p}}   }{2^{5+\frac{1}{3}}\pi ^\frac{4}{3}\alpha^\frac{1}{3}   }\left ( \frac{\phi }{m_{\rm{p}}}  \right )^\frac{4n-4}{3}.
\end{eqnarray}
The requirement for the potential to be real is
\begin{eqnarray}
\label{311}
\left(\frac{\phi}{m_{\mathrm{p}}}\right)^{2n} \geq-\frac{n^\frac{4}{3}   }{2^{2+\frac{1}{3}}\pi ^\frac{1}{3}\alpha^\frac{1}{3}\beta^\frac{2}{3}   }\left ( \frac{\phi }{m_{\rm{p}}}  \right )^\frac{4n-4}{3},
\end{eqnarray}
This condition depends on the parameters $n$ and $\alpha$, which can only be determined through observations. Inflation concludes when the acceleration $\ddot{a}$ becomes zero. This is indicated by Eq. \eqref{slowend}, suggesting that the scalar field at the end of inflation could satisfy
\begin{eqnarray}
\sqrt{2 c_{s}}{\mathcal{H}_{1} n \left(\frac{\phi }{m_{\rm{p}}}\right)^{n-1}}=\mathcal{H}_{1} \left(\frac{\phi }{m_{\rm{p}}}\right)^n.
\end{eqnarray}
Solving this equation yields
\begin{eqnarray}
\label{18}
\phi_{\rm{e}}=\frac{\sqrt{2} m_{\rm{p}} n}{\sqrt[4]{3}}.
\end{eqnarray}
By employing the relation for the number of e-folds, as given in \eqref{efold}, the scalar field at the onset of inflation can be determined as follows:
\begin{eqnarray}
\label{181}
\phi_{\rm{i}}=54^{-\frac{1}{2n+4} } m_{\rm{p}}\left [ 2^\frac{n+3}{3}3^\frac{4-n}{6}n^\frac{2n+4}{3} -\beta^{-\frac{2}{3} } \left ( n+2 \right )N\left ( -\frac{n}{\pi \alpha}  \right ) ^\frac{1}{3} \right ]^\frac{3}{2n+4}   .
\end{eqnarray}
Utilizing the Friedmann equation (\ref{9}) and the Hubble parameter (\ref{pow}), the scalar and tensor power spectra in the slow-roll regime can be derived from Eqs. (\ref{ps}) and (\ref{pt}) as follows \cite{Garriga:1999vw,Unnikrishnan:2012zu}:
\begin{eqnarray}
\label{ps1}
\mathcal{P}_S=2^\frac{2}{n+2}3^{\frac{-7n-2}{2n+4}}\alpha^\frac{1}{3}\beta^\frac{8}{3}     \pi ^\frac{-2}{3}n^\frac{-4}{3}   \left[2^\frac{n+3}{3}3^\frac{4-n}{6}n^\frac{2n+4}{3}-(n+2) N\left(-\frac{n}{\pi \alpha}\right)^\frac{1}{3}  \beta^\frac{-2}{3}  \right]^\frac{4n+2}{n+2}   ,
\end{eqnarray}
and
\begin{eqnarray}
\label{pt1}
\mathcal{P}_T= 2^{\frac{3 n+8}{n+2}} 3^{\frac{-6n}{2n+4}}  \beta^{2}  \pi^{-1}  \left[2^\frac{n+3}{3}3^\frac{4-n}{6}n^\frac{2n+4}{3}-(n+2) N\left(-\frac{n}{\pi \alpha} \right)^\frac{1}{3}\beta^\frac{-2}{3} \right]^\frac{3n}{n+2} .
\end{eqnarray}
\begin{table}
\centering
\renewcommand\arraystretch{4}
\setlength{\tabcolsep}{4mm}{
\begin{tabular}{c|c c}
\hline
  & $\alpha$   & $\beta$   \\
\hline
$N=55$ & $2.266\times 10^{17}$   & $2.324\times 10^{-6}$   \\ \cline{1-3}
$N=60$ & $3.124\times10^{17}$ & $2.525\times10^{-6}$  \\ \cline{1-3}
$N=65$ & $3.823\times10^{17}$ & $2.814\times10^{-6}$  \\
\hline
\end{tabular}}
\caption{The parameters $\alpha$ and $\beta$ are determined, utilizing $\mathcal{P}_{S}\simeq 2\times 10^{-9}$ from the 2015 Planck data \cite{Planck:2015sxf}, alongside $n=-0.07$, $r=0.02$, and $n_{\rm{s}}=0.9668$ predicted by the 2018 Planck data \cite{Planck:2018jri}.
}
\label{ab}
\end{table}
\begin{figure}
\includegraphics[height=8cm,width=10cm]{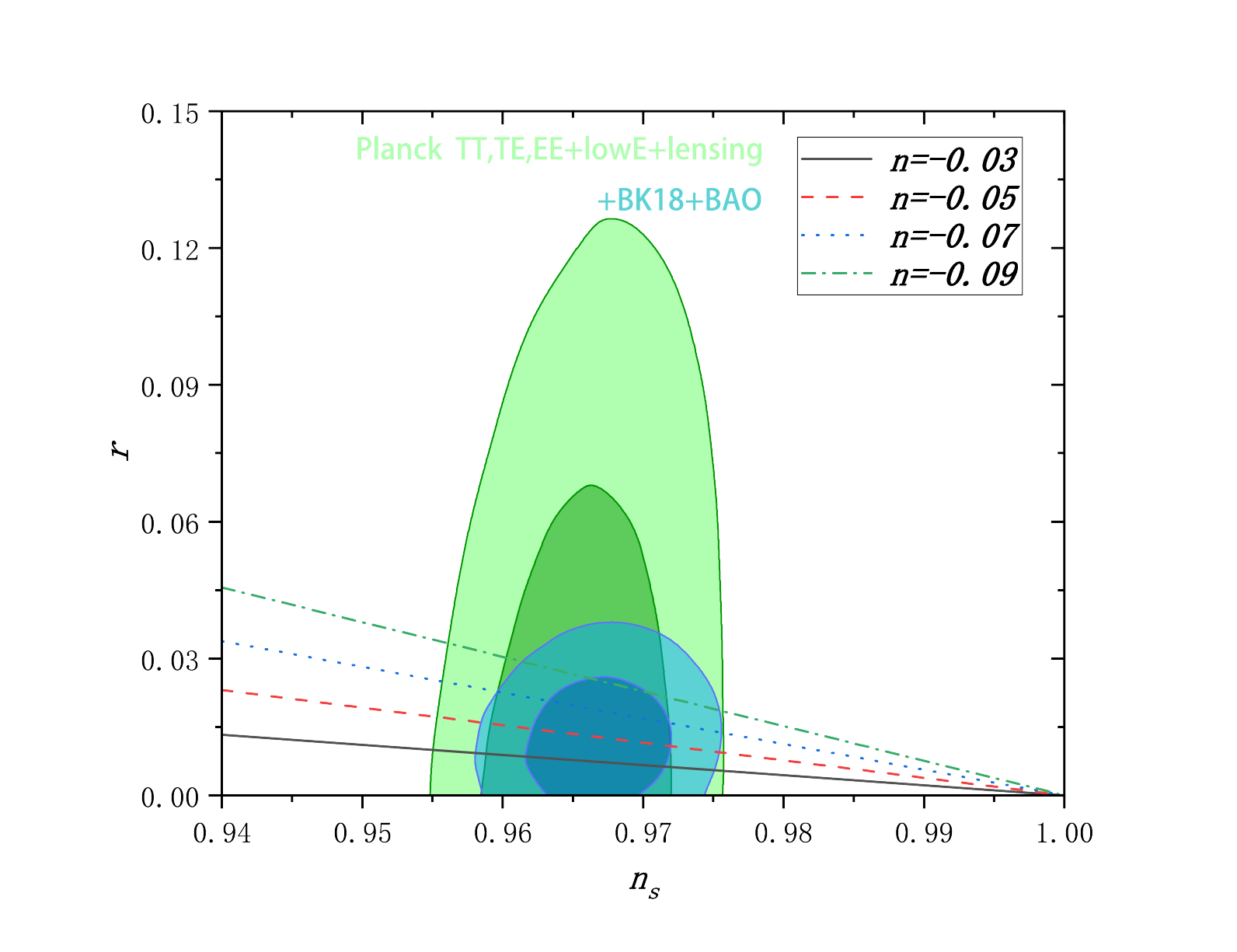}
\caption{The plot depicts the tensor-to-scalar ratio $r$ against the scalar spectral index $n_{\rm{s}}$, considering values of $n$ as -0.03, -0.05, -0.07, and -0.07, respectively. The marginalized joint 68\% and 95\% confidence level (C.L.) regions for $r$ and $n_{\rm{s}}$ at $k=0.002$ Mpc$^{-1}$, derived from the Planck 2018 data \cite{Planck:2018jri}, are presented in blue and green, respectively.}
\label{rns}
\end{figure}
\begin{figure}
\centering
\includegraphics[height=8cm,width=10cm]{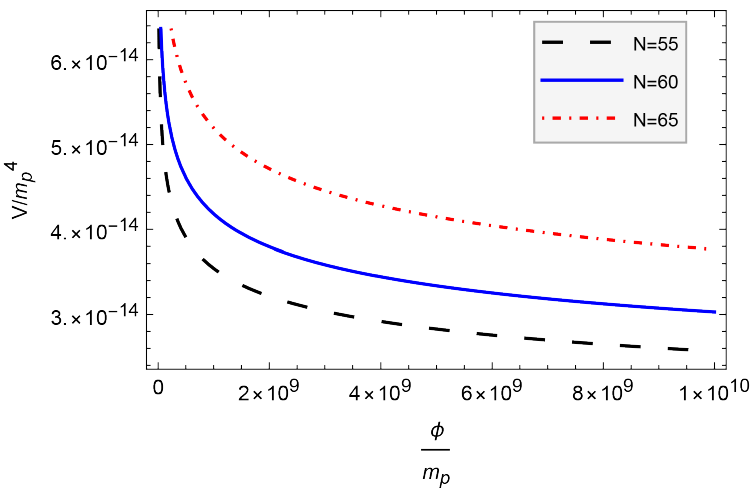}
\caption{The potential $V$ is depicted as a function of the k-field, with parameter values as specified in Table \ref{ab}.}
\label{fig1}
\end{figure}
Given the nearly constant nature of $H$ during slow-roll inflation and the constant speed of sound in our model, it follows that at sound horizon exit: $d/d\ln k\simeq d/dN$. This equivalence arises directly from the amplitude of scalar and tensor perturbations, allowing for straightforward derivation of the scalar and tensor spectral indices as:
\begin{eqnarray}
\label{ns1}
n_{\rm{s}}\equiv 1+\frac{d\ln \mathcal{P}_S}{d\ln k}=1-\frac{(4n+2)\left(-\frac{n}{\pi \alpha}\right) ^{\frac{1}{3} }}{(n+2)  N\left(-\frac{n}{\pi \alpha} \right)^\frac{1}{3}- 2^\frac{n+3}{3} 3^\frac{4-n}{6}  \beta^\frac{2}{3}n^\frac{2n+4}{3}  },
\end{eqnarray}
and
\begin{eqnarray}
\label{nt}
n_{\rm{t}}\equiv \frac{d\ln\mathcal{P}_T }{d\ln k}=\frac{3n\left(-\frac{n}{\pi \alpha} \right)^\frac{1}{3} }{2^\frac{n+3}{3} 3^\frac{4-n}{6}\beta^\frac{2}{3}n^\frac{2n+4}{3}-(n+2)N(-\frac{n}{\pi \alpha} )^\frac{1}{3}}.
\end{eqnarray}
The tensor-to-scalar ratio is
\begin{eqnarray}
\label{r}
r\equiv\frac{\mathcal{P}_T}{\mathcal{P}_S}= \frac{3^\frac{1}{2}8n^\frac{4}{3}  }{\pi ^\frac{1}{3}\alpha^\frac{1}{3}\left [2^\frac{n+3}{3}3^\frac{4-n}{6} \beta^\frac{2}{3}n^\frac{2n+4}{3}-\left ( n+2 \right ) N\left ( -\frac{n}{\pi \alpha}  \right ) ^\frac{1}{3}   \right ]  }.
\end{eqnarray}
We note that when evaluating $r$ the expression for the scalar power spectrum $\mathcal{P}_S$(k) in \eqref{ps} is evaluated at sound horizon exit ($aH=c_sk)$ while the
corresponding tensor power spectrum $\mathcal{P}_T$(k) in \eqref{pt} is evaluated at horizon exit ($aH = k$). However, since the speed of sound $c_s$ does not depend upon time, and $H$ is almost constant during slow roll, the value of the field $\phi$ at sound horizon exit is approximately the same as at horizon exit \cite{Garriga:1999vw,Unnikrishnan:2012zu}.

Upon examination of Eqs. \eqref{nt} and \eqref{r}, it is evident that the consistency relation for k-inflation, $r=-8c_{\rm{s}}n_{\rm{t}}$, holds \cite{Garriga:1999vw}. Combining Eqs. (\ref{ns1}) and (\ref{r}) results in
\begin{eqnarray}
\label{consis}
r=\left ( n_{\rm s} -1 \right ) \frac{4\times3^\frac{1}{2} n}{2n+1}.
\end{eqnarray}
Constrained by Planck TT, TE, EE+lowE+lensing+BK15+BAO data \cite{Planck:2018jri}, the scalar spectral index is approximately $n_{\rm{s}}= 0.9668\pm0.0037$, and the tensor-to-scalar ratio $r$ is constrained as $r<0.063$. Under this limitation, if taking $r=0.02$ and $n_{\rm{s}}=0.9668$, we deduce $n=-0.07$ from Eq. (\ref{consis}). By substituting these values into (\ref{ns1}) and considering $N=55$, $N=60$, and $N=65$, we can solve for $\alpha$ as functions of $\beta$. Further, by inserting these values and utilizing $\mathcal{P}_S\simeq 2\times 10^{-9}$ \cite{Planck:2015sxf} into (\ref{ps}), we obtain the parameter values for $\alpha$ and $\beta$, as summarized in Table \ref{ab}.

With Eq. (\ref{consis}) and treating the scalar spectral index $n_{\rm{s}}$ as a variable, we present a plot of the tensor-to-scalar ratio $r$ as a function of $n_{\rm{s}}$ in Fig. \ref{rns} with $n=-0.03$, $n=-0.05$, $n=-0.07$, and $n=-0.09$, respectively. The marginalized joint 68\% and 95\% confidence level (C.L.) regions for $r$ and $n_{\rm{s}}$ at $k=0.002$ Mpc$^{-1}$ are derived from Planck 2018 data \cite{Planck:2018jri}, displayed in blue and green, respectively. We observe that for $n\geq -0.9$ the model exhibits good agreement with observational constraints 68\% C.L..

Upon inserting $n=-0.07$ and the values of $\alpha$ and $\beta$ from Table \ref{ab} into Eq. (\ref{potential}), we generate a plot illustrating the scalar potential $V$ as a function of the k-field in Fig. \ref{fig1}. It is evident that the potential exhibits a decreasing trend and approaches zero as the scalar field increases.

Finally, the running of the scalar spectral index $n_{\rm{run}}$ is found to be
\begin{eqnarray}
n_{\rm{run}}\equiv \frac{d\,n_{\rm s}}{d\,\ln k}=-\frac{2 \times 3^{\frac{n}{3}} (-\pi)^{-\frac{2}{3}} (n+2) (2 n+1)}{\left[3^{\frac{n}{6}}(-\pi)^{-\frac{3}{2}}  (n+2) N-3^{\frac{2}{3}} 2^{\frac{n}{3}+1} \sqrt[3]{\alpha } \beta ^{\frac{2}{3}} n^{\frac{2 n}{3}+1}\right]^2}.
\end{eqnarray}
\begin{figure}
\centering
\includegraphics[height=8cm,width=10cm]{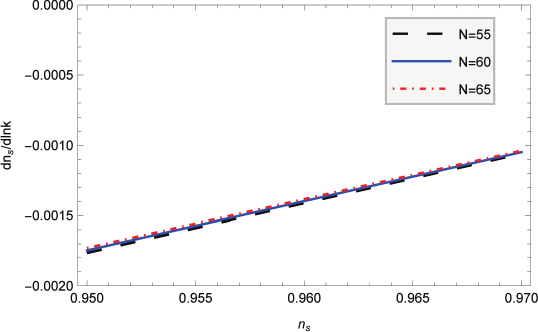}
\caption{The plot shows the running of the scalar spectral index $dn_{\rm s} /d \ln k$ versus the scalar spectral index $n_{\rm s}$ with three different values for the number of e-folds: the black, blue, and red lines correspond to $N$ = 55, 60, and 65, respectively.}
\label{fig3}
\end{figure}
\begin{figure}
\centering
\includegraphics[height=8cm,width=10cm]{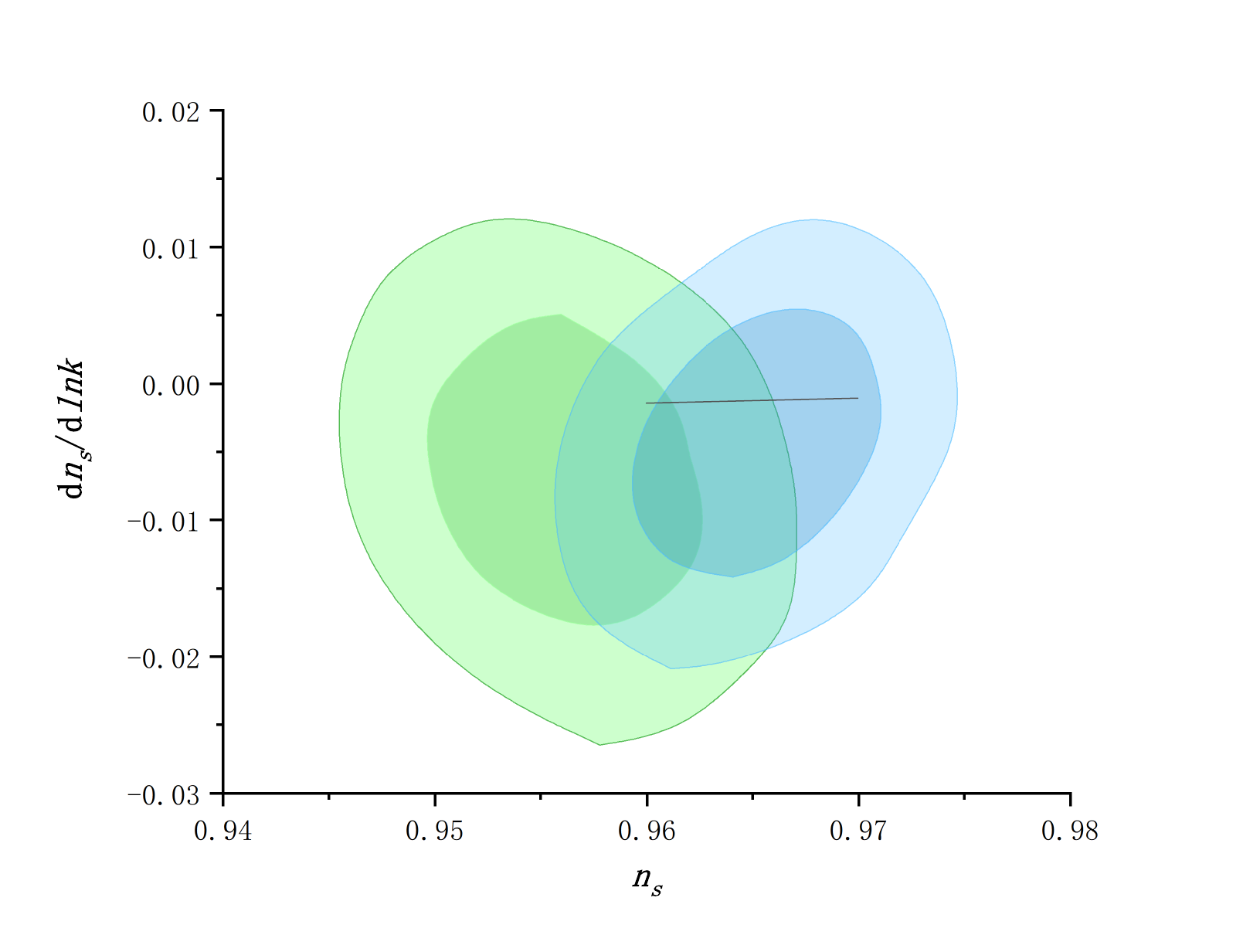}
\caption{The plot shows the two-dimensional marginalized joint confidence contours for
($n_{\rm s}$, $dn_{\rm s} /d \ln k$), at the 68 and 95\% C. L., in the presence of a non-zero
tensor contribution, from the Planck 2015 data (green) \cite{Planck:2015sxf} as well as the Planck 2018 data (blue) \cite{Planck:2018vyg}. Additionally, we depict the predictions of our scenario with $N=60$ (black-solid line).}
\label{fig4}
\end{figure}
We have obtained three different curves by fixing the number of e-folds to $N$ = 55, 60, and 65, which are shown in Fig. \ref{fig3}. The difference between these three curves is very small. Within the allowable range of observation values for $n_{\rm s}$, the values of $dn_{\rm s} /d \ln k$ also change very little.

In order to compare the previous predictions with the observational data, in Fig. \ref{fig4} we show the two-dimensional marginalized joint confidence contours for ($n_{\rm{s}}$, $dn_{\rm{s}} /d \ln k$) at the 68\% and 95\% CL in the presence of a non-zero tensor contribution from the Planck 2015 data as well as the Planck 2018 data. We also depict the predictions of our scenario with the e-folding value $N$ being 60, which clearly seems to be in perfect agreement with observational data.

\section{conclusion}

We have explored a particular form of k-inflation utilizing the Hamilton-Jacobi approach. After deriving the general equation characterizing the model, we posited that the Hubble parameter could be expressed as a power-law function of the k-field. The observables of the model, including the scalar power spectrum, the tensor-to-scalar ratio, the scalar and tensor spectral indices, have been derived. The model's parameters were then constrained using Planck data, and the specific form of the potential was presented. By visualizing the potential's behavior as the k-field increases, we demonstrated that it exhibits a decreasing trend. Furthermore, we plotted the tensor-to-scalar ratio $r$ and the the running of the scalar spectral index $n_{\rm{run}}$ against the scalar spectral index $n_{\rm{s}}$ for various values of the parameter $n$, illustrating that the model aligns well with observational data.

Our treatment here should be straightforward to generalize to other scalar field inflationary models. In standard inflation, the consistency relation is $r=-8n_{\rm t}$ \cite{Liddle:1993fq}, which is modified as $r=-8c_{\rm s}n_{\rm t}$ in k-inflation \cite{Garriga:1999vw}. This relation can be used as test in the future for the model we proposed here. It is also worth studying whether can we use the Cosmic Microwave Background temperature, polarization data from Planck 2018 data release and updated likelihoods to constrain the model proposed here \cite{Hazra:2021eqk,Montandon:2020kuk,Antony:2021bgp}.

\begin{acknowledgments}
We thank Xinyi Zhang for helpful discussions. This study is supported in part by National Natural Science Foundation of China (Grant No. 12333008) and Hebei Provincial Natural Science Foundation of China (Grant No. A2021201034).

\textbf{Data Availability Statement}: No Data associated in the manuscript.

\end{acknowledgments}

\bibliographystyle{ieeetr}
\bibliography{X322}
\end{document}